\documentclass[%
 aip,
rsi,%
 amsmath,amssymb,
 reprint,%
]{revtex4-1}

\usepackage{amsmath}
\usepackage{amssymb}
\usepackage{amsfonts}
\usepackage{amssymb}
\usepackage[pdftex]{graphicx}
\usepackage{subfigure} 
\usepackage[english]{babel}
\usepackage{braket}
\usepackage{color}
\usepackage{mathtools}
\usepackage{threeparttable}
\usepackage[colorlinks,linkcolor=blue,anchorcolor=blue,citecolor=blue,urlcolor=blue]{hyperref}
\usepackage{babel}
\makeatother

\begin{document}

\preprint{APS/123-QED}

\title{Probabilistic vortex crossing criterion for superconducting nanowire single-photon detectors}

\author{Saman Jahani}
\affiliation{%
School of Electrical and Computer Engineering and Birck Nanotechnology Center, Purdue University, West Lafayette, IN 47907 USA.
}%
\affiliation{%
Current Address: Moore Laboratory, Department of Electrical Engineering, California Institute of Technology, Pasadena, CA 91125, USA.
}%
\author{Li-Ping Yang}%
\affiliation{%
School of Electrical and Computer Engineering and Birck Nanotechnology Center, Purdue University, West Lafayette, IN 47907 USA.
}%
\author{Adri{\'a}n Buganza Tepole}%
\affiliation{%
School of Mechanical Engineering,
Purdue University, West Lafayette, IN 47907 USA.
}%
\author{Joseph C. Bardin}%
\affiliation{%
Department of Electrical and Computer Engineering,
University of Massachusetts at Amherst, Amherst, MA 01003 USA.
}%
\author{Hong X. Tang}%
\affiliation{%
Department of Electrical Engineering, Yale University, New Haven, CT 06511, USA.}%
\author{Zubin Jacob}%
\email{zjacob@purdue.edu}
\affiliation{%
School of Electrical and Computer Engineering and Birck Nanotechnology Center, Purdue University, West Lafayette, IN 47907 USA.
}%


\begin{abstract}
Superconducting nanowire single-photon detectors have emerged as a promising technology for quantum metrology from the mid-infrared to ultra-violet frequencies. Despite the recent experimental successes, a predictive model to describe the detection event in these detectors is needed to optimize the detection metrics.  
Here, we propose a probabilistic criterion for single-photon detection based on single-vortex (flux quanta) crossing the width of the nanowire. Our model makes a connection between the dark-counts and photon-counts near the detection threshold.
The finite-difference calculations demonstrate that a change in the bias current distribution as a result of the photon absorption significantly increases the probability of single-vortex crossing even if the vortex potential barrier has not vanished completely. We estimate the instrument response function and show that the timing uncertainty of this vortex tunneling process corresponds to a fundamental limit in timing jitter of the click event. We demonstrate a trade-space between this intrinsic (quantum) timing jitter, quantum efficiency, and dark count rate in TaN, WSi, and NbN superconducting nanowires at different experimental conditions. 
Our detection model can also explain the experimental observation of exponential decrease in the quantum efficiency of SNSPDs at lower energies. This leads to a pulse-width dependency in the quantum efficiency, and it can be further used as an experimental test to compare across different detection models.
\end{abstract}

\maketitle


\section{\label{sec:level1}Introduction}

Advancements in quantum technologies strongly depends on improvement in the detection of light at the single-photon level. This requires near-unity quantum efficiency, sub-picosecond timing uncertainty (timing jitter), sub-milihertz dark count rate, large bandwidth, and fast reset time \cite{hadfield_single-photon_2009}. Superconducting nanowire single photon detectors (SSPDs or SNSPDs) are highly promising detectors in a broad range of frequencies from mid-infrared to ultraviolet \cite{goltsman_picosecond_2001, natarajan_superconducting_2012, takesue_quantum_2007, eisaman_invited_2011,zhao_single-photon_2017,holzmansuperconducting} with near unity quantum efficiency \cite{marsili_detecting_2013}, picosecond-scale timing jitter \cite{korzh_demonstrating_2018, korzh2018wsi, sidorova_intrinsic_2018, pernice_high-speed_2012}, fast reset time \cite{tarkhov_ultrafast_2008}, and milihertz dark count rate \cite{schuck_waveguide_2013, charaev2016current}. They are composed of a thin superconducting nanowire which is biased slightly below the superconducting critical current. Photon absorption triggers a phase transition giving rise to generation of a voltage pulse which is measured by a readout circuit connected to the nanowire. 

Owing to the experimental progress on reducing the amplification noises and the uncertainty of the photon absorption location, recent breakthrough results have shown timing jitter below 10 ps \cite{korzh_demonstrating_2018, korzh2018wsi, sidorova_intrinsic_2018, zhu2018superconducting, sidorova2018timing}. Hence, the response function of SNSPDs to a single-photon has approached its intrinsic response limit which only depends on the microscopic mechanism of light-matter interaction in nanowires. To further improve the performance of these detectors, it is required to understand the microscopic mechanism and the trade-space of the photon detection event in these detectors. 

Over the past two decades, several important detection models have been proposed to explain the microscopic mechanism of the formation of the first resistive region in SNSPDs \cite{goltsman_picosecond_2001, natarajan_superconducting_2012,epstein_vortex-antivortex_1981,renema_experimental_2014, bulaevskii_vortex-assisted_2012, engel_numerical_2013, engel2015detection, vodolazov2017single}. In the simplest model, it is assumed that the energy of the absorbed photon increases the temperature at the absorption site leading to the nucleation of a hot-spot which causes the current to be directed to the sides \cite{goltsman_picosecond_2001, marsili_hotspot_2016, korneeva_comparison_2017}. This may cause the current at the edge to surpass the critical depairing current causing formation of a normal conducting region across the width. In another model, the depletion of superconducting electrons around the absorption site is responsible for the formation of the resistive region \cite{semenov_spectral_2005}. Recently, some models have suggested that the motion of vortices or vortex-antivortex pairs can also induce a phase transition at a lower applied bias current \cite{epstein_vortex-antivortex_1981,zotova_photon_2012, bulaevskii_vortex-induced_2011, engel_numerical_2013, vodolazov2015vortex, vodolazov2017single}.  
Although each of these models explain most of the macroscopic behaviors of SNSPDs, existing models cannot explain or predict the trade-off between the quantum efficiency, timing jitter, and dark counts and their fundamental limits in SNSPDs.

\begin{figure}[ht]
\centering
\begin{tabular}{cc}
\includegraphics[width=8.5cm]{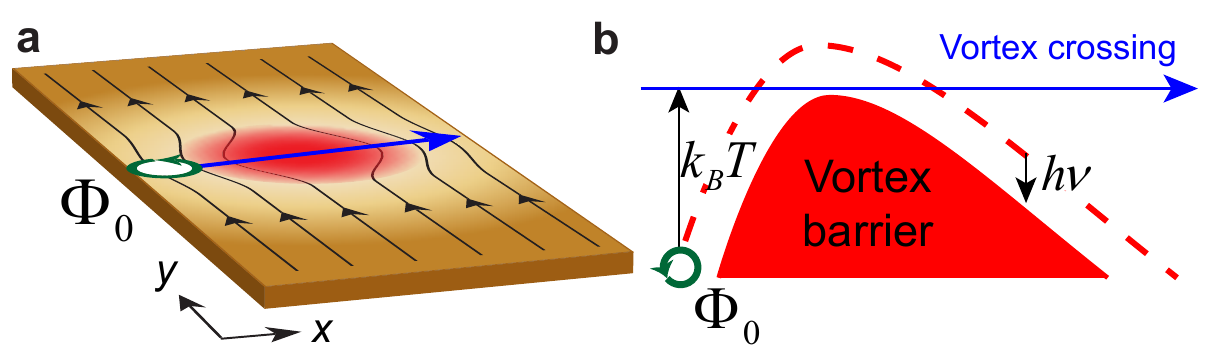}
\end{tabular}
\caption{ {\bf Superconducting nanowire single photon detectors (SNSPDs)}. {\bf a,} When a photon falls on the detector, quasi-particles (QPs) are generated and the bias current is redistributed. Vortices with magnetic flux quantum of ${\Phi}_{0}$ are the topological defects of a thin superconductor and are nucleated at the nanowire edge. They can move to the other edge due to the force exerted by the bias current. {\bf b,} Before the photon absorption, the vortex potential barrier does not allow them to move easily. However, due to the QP multiplication and current redistribution, the barrier reduces and vortices which are thermally excited can escape the barrier and cross the width of the nanowire. This process generates a voltage pulse propagating to the two ends of the detector. We provide a probabilistic click definition using this detection event.
}
\label{fig:SSPD_Schematic}
\end{figure}

In this paper, we construct a connection between the photon-induced counts and the dark counts in SNSPDs, which has been recently observed experimentally around the detection current \cite{korneeva2018optical, knehr2019nanowire}. We propose a probabilistic criterion for single-photon detection corresponding to the single-vortex crossing from one edge of the nanowire to the other edge. First, we numerically calculate the time-dependent current distribution after the photon absorption and its effect on the vortex potential. We propose that due to the change in the distribution of the superconducting electrons, the probability of the vortex crossing is significantly enhanced even if the vortex potential barrier has not vanished completely. 

Then, we define the detection probability based on the probability of the single-vortex crossing, because the energy released by one vortex moving across the width is enough to induce a phase transition in the superconducting nanowire. We show that the probabilistic behavior of the single-vortex crossing results in an intrinsic timing jitter on the click event. This intrinsic quantum timing jitter cannot be eliminated even if the geometric position of photon absorption is known, however, it can be reduced by engineering the structure and the experimental conditions at the cost of a degradation of the quantum efficiency and/or an increase in the dark count rate. 

Finally, we calculate the quantum efficiency spectrum and show that the quantum efficiency does not suddenly drop to zero when the photon energy is below a threshold. We propose that the response of the detector to the photon pulse-width can be different for the various detection models. Moreover, the quantum efficiency predicted by our model is strongly dependent on the pulse-width. This effect has not been predicted by the previous detection models. Our work unifies previously known ideas of vortex crossing phenomenon with the POVM approach of quantum optics to propose a probabilistic detection criterion for SNSPDs. We propose some observable quantities which can be used to experimentally verify the validity of our probabilistic model.
Our model focuses around the detection threshold (quantum-efficiency$\approx$1) where photons do not have enough energy to form a normal conducting hot-spot and the probabilistic behavior of vortices is more significant. This is not in contradiction to observations of the vortex/anti-vortex pair unbinding. For higher energies or higher bias currents, the formation of a hot-spot and, as a result, vortex/anti-vortex pair unbinding might happen before the probabilistic tunneling of a single-vortex from the edges \cite{vodolazov2017single, allmaras2019intrinsic}.

\section{Detection mechanism}

Detection mechanism in SNSPDs consists of three steps: (a) photon absorption and breaking the superconducting electron pairs (known as Cooper pairs) to quasi-particles (QPs) leading to formation of a hot-spot; (b) as a result of the depletion of the Cooper pairs, the superconducting order parameter is suppressed. This causes the current density at the absorption location to be reduced and directed to the sides as illustrated in Fig.~\ref{fig:SSPD_Schematic}a; (c) the change in the Cooper pairs and current density reduces the vortex potential barrier and vortices can move across the nanowire and release a measurable voltage pulse (Fig.~\ref{fig:SSPD_Schematic}b).

\begin{figure}[ht]
\centering
\begin{tabular}{cc}
\includegraphics[width=8.5cm]{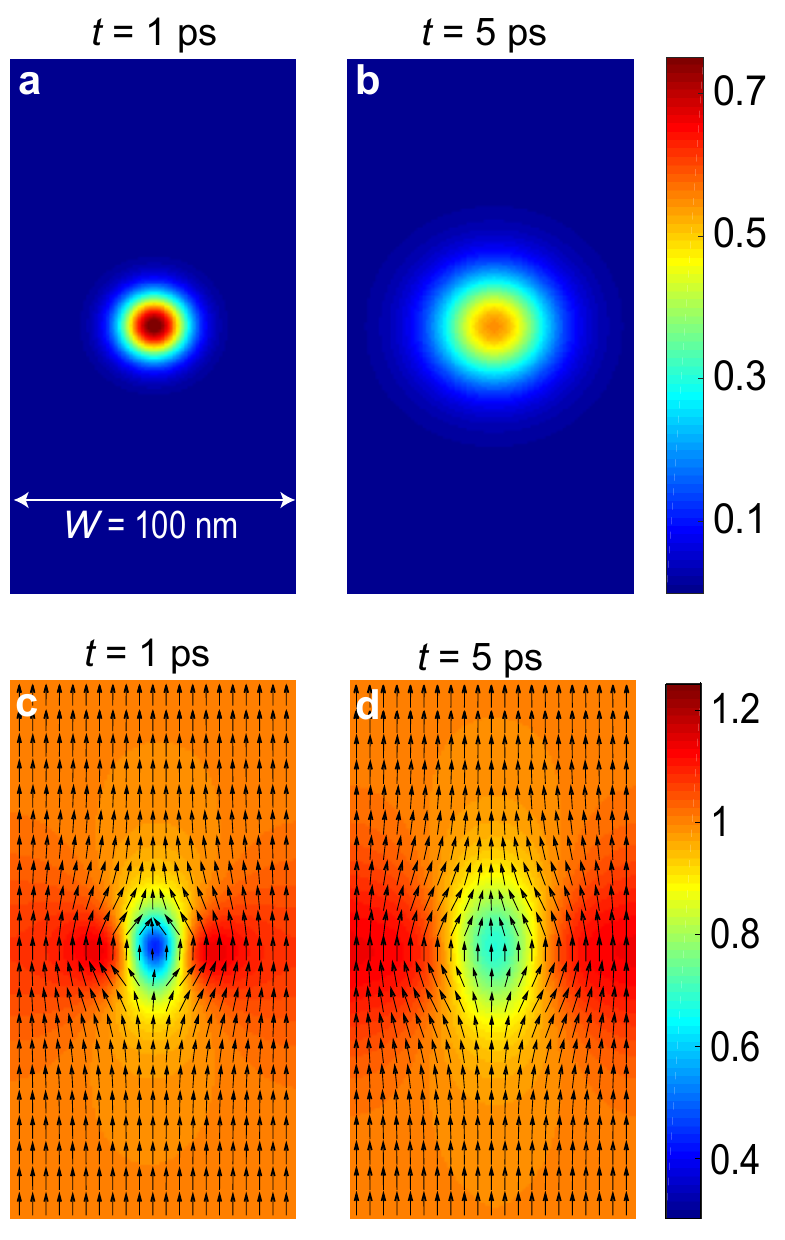}
\end{tabular}
\caption{{\bf a,} and {\bf b,} QP distribution, $C_{qp}(\vec{r},t)$, normalized to the initial density of superconducting electrons, $n_{se,0}$, in a TaN SNSPD at $t=1$~ps and $t=5$~ps, respectively. It is assumed the photon is absorbed at $t=0$. $T=0.6$~$^{\circ}$K. The photon energy is $h\nu=1.5$~eV. The nanowire width, length, and thickness are 100 nm, 1000 nm, and 5 nm, respectively. {\bf c,} and {\bf d,}  Normalized Current density in the $y$ direction at $t=1$~ps and $t=5$~ps, respectively. The current density is normalized to the applied bias current. The arrows represent the current density vector, $\vec{j}(\vec{r},t)$. Due to the hot-spot formation, current is redistributed and directed to the side walls.  
}
\label{fig:SSPD_distribution}
\end{figure}

\begin{figure}[ht]
\centering
\begin{tabular}{cc}
\includegraphics[width=8.5cm]{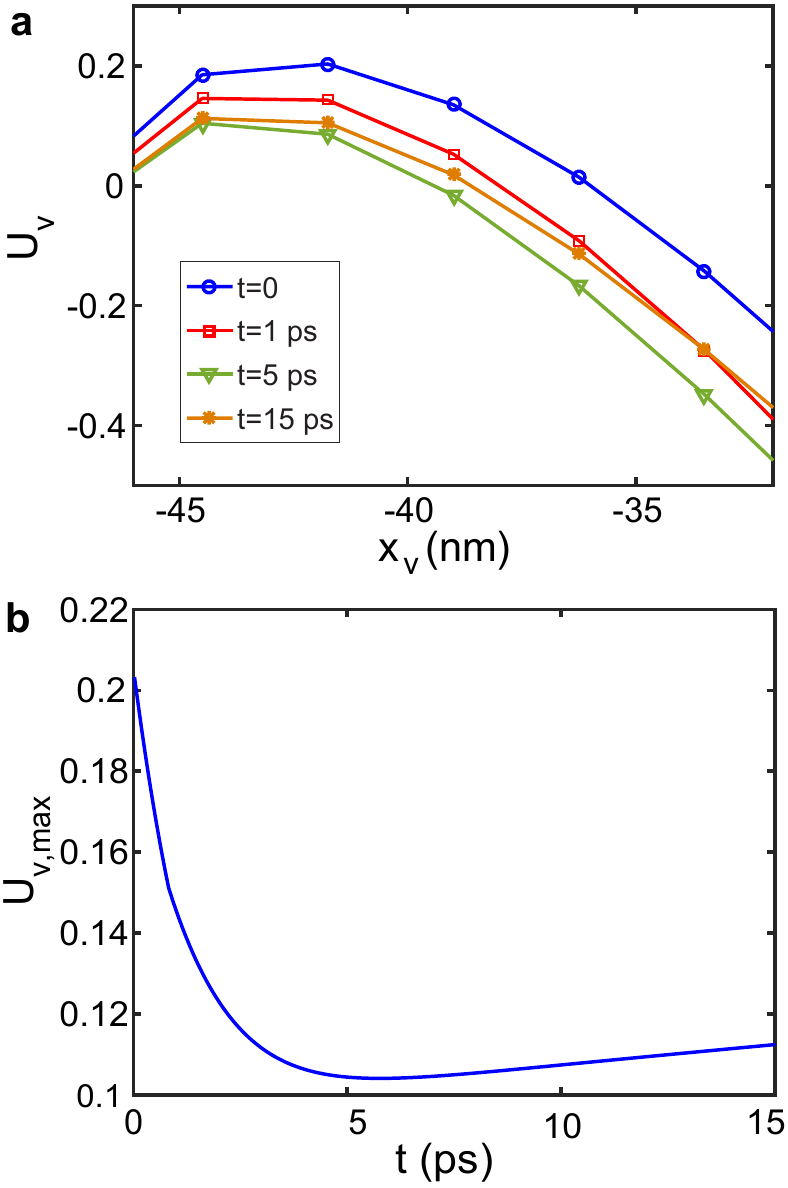}
\end{tabular}
\caption{ {\bf Vortex barrier dynamics} {\bf a,} Vortex potential as a function of the vortex location, $x_v$, around the saddle point after the photon absorption for a TaN SNSPD. $T=0.6$~$^{\circ}$K, $W=100$~nm, and $h\nu=1.5$~eV. {\bf b,} Potential barrier peak as a function of time. A change in the Cooper pairs density and current distribution reduce the barrier hight. The potential has been normalized to the characteristic vortex energy, ${\varepsilon}_0$.
}
\label{fig:SSPD_Potential}
\end{figure}

These three steps have been quantitatively described in the appendix. Our finite-difference calculations of QPs distribution based on the diffusion model \cite{engel_numerical_2013} for a TaN SNSPD is illustrated in Figs.~\ref{fig:SSPD_distribution}a and \ref{fig:SSPD_distribution}b at $t=1$ ps and $t=5$ ps, respectively. We assume a photon with the energy of $h\nu=1.5$~eV falls at the center of the SNSPD at $t=0$. The width and the length of the nanowire are $W=100$~nm and $L=1000$~nm, respectively. Figures~\ref{fig:SSPD_distribution}c and \ref{fig:SSPD_distribution}d display the numerical calculation of the current density normalized to the bias current. It is seen that due to the hot-spot formation at the center, the current is directed to the side-walls of the nanowire, and as a result, the vortex potential barrier is reduced as shown in Fig.~\ref{fig:SSPD_Potential}. If the bias current or the photon energy are high enough, the potential barrier can be vanished completely.

After the single-photon transduction, several processes compete with each other to form the initial normal conducting cross-section. Depending on which one occurs first, different detection models have been proposed. In hot-spot model, it is assumed that the formation of hot-spot is responsible for the phase transition \cite{goltsman_picosecond_2001, korneeva_comparison_2017}. Nucleation of the hot-spot causes the bias current to be directed to the side-walls. If the current density at the edge surpasses the despairing critical current ($I_{\rm edge}\geq I_{c,dep}$), it induces a phase transition to the normal conducting state at the edge and the normal conducting region expands across the width.

In QP model, there is no need to destroy the superconductivity by surpassing the critical current \cite{semenov_spectral_2005}. If the Cooper pairs are depleted inside a volume with a thickness of at least one coherence length ($\xi$-slab), the phase coherence is destroyed which results in a phase transition. This requires the number of QPs inside the $\xi$-slab ($N_{QP}^{slab}$) to exceed the number of the superconducting electrons inside the $\xi$-slab: $N_{QP}^{slab}/N_{se}^{slab}\geq 1-I_b/I_{c,dep}$, where $N_{se}^{slab}$ is the initial superconducting-electron number inside the slab before applying the bias current ($I_b$).

Vortices can also be responsible for the trigger of a single-photon induced phase transition in SNSPDs. If the photon transduction causes the vortex potential barrier ($U_v$) to vanish, vortices move across the width and induce a phase transition \cite{engel_numerical_2013,bulaevskii_vortex-assisted_2012}. In the next section, we show that even if the barrier has not completely vanished and the kinetic energy of the vortices is not enough to surmount the barrier, there is a considerable probability of single-vortex crossing. This quantum tunneling process causes a new source of timing jitter for the detection event.

\begin{figure}[htbp]
\centering
\begin{tabular}{cc}
\includegraphics[width=8cm]{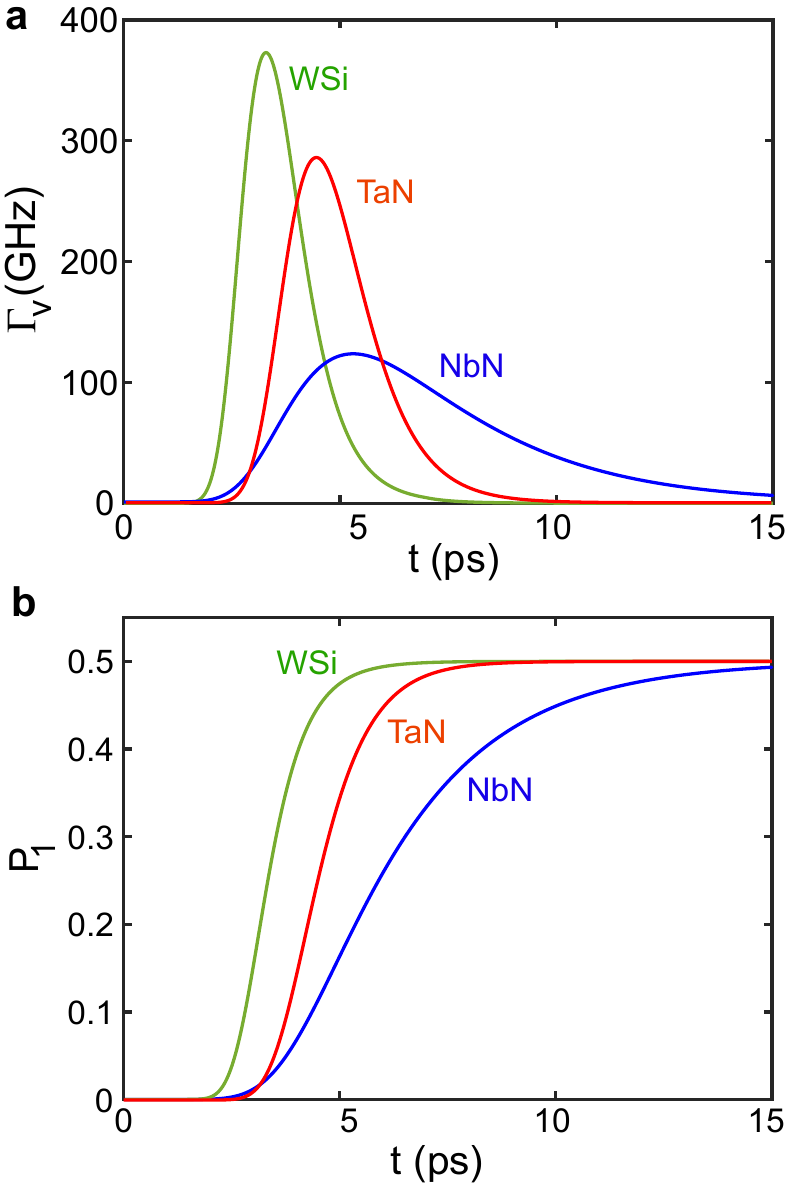}
\end{tabular}
\caption{ {\bf Vortex crossing rate and probability.} {\bf a,} Single-vortex crossing rate as a function of time for NbN, TaN, and WSi SNSPDs. The bias current has been set to achieve a single photon detection probability of 0.5 for a photon energy of $h\nu=1.5$~eV, which is $0.96I_{\rm SW}$, $0.93I_{\rm SW}$, and $0.65I_{\rm SW}$, for NbN, TaN and WSi SNSPDs, respectively. The reduced temperature ($T/T_c$) for TaN, NbN, and WSi at $T=0.6$~$^{\circ}$K are 0.07, 0.05, and 0.15. $W=100$~nm. $I_{\rm SW}$ is defined as the minimum bias current at which the detector clicks in the time-bin of the single-photon arrival even if the photon is not absorbed. Material parameters are derived from ref. \cite{engel_detection_2015}. The enhancement in the vortex crossing rate is as a result of the suppression of the potential barrier. The probability of vortex crossing at the maximum rate is higher. However, there is considerable uncertainty in the vortex crossing time which results in a timing jitter in detection event. {\bf b,} The evolution of vortex crossing probability as the vortex crossing rate changes. There is a steep change in the probability as the crossing rate goes up. 
}
\label{fig:SSPD_Gamma}
\end{figure}

\begin{figure}[htbp]
\centering
\begin{tabular}{cc}
\includegraphics[width=8.5cm]{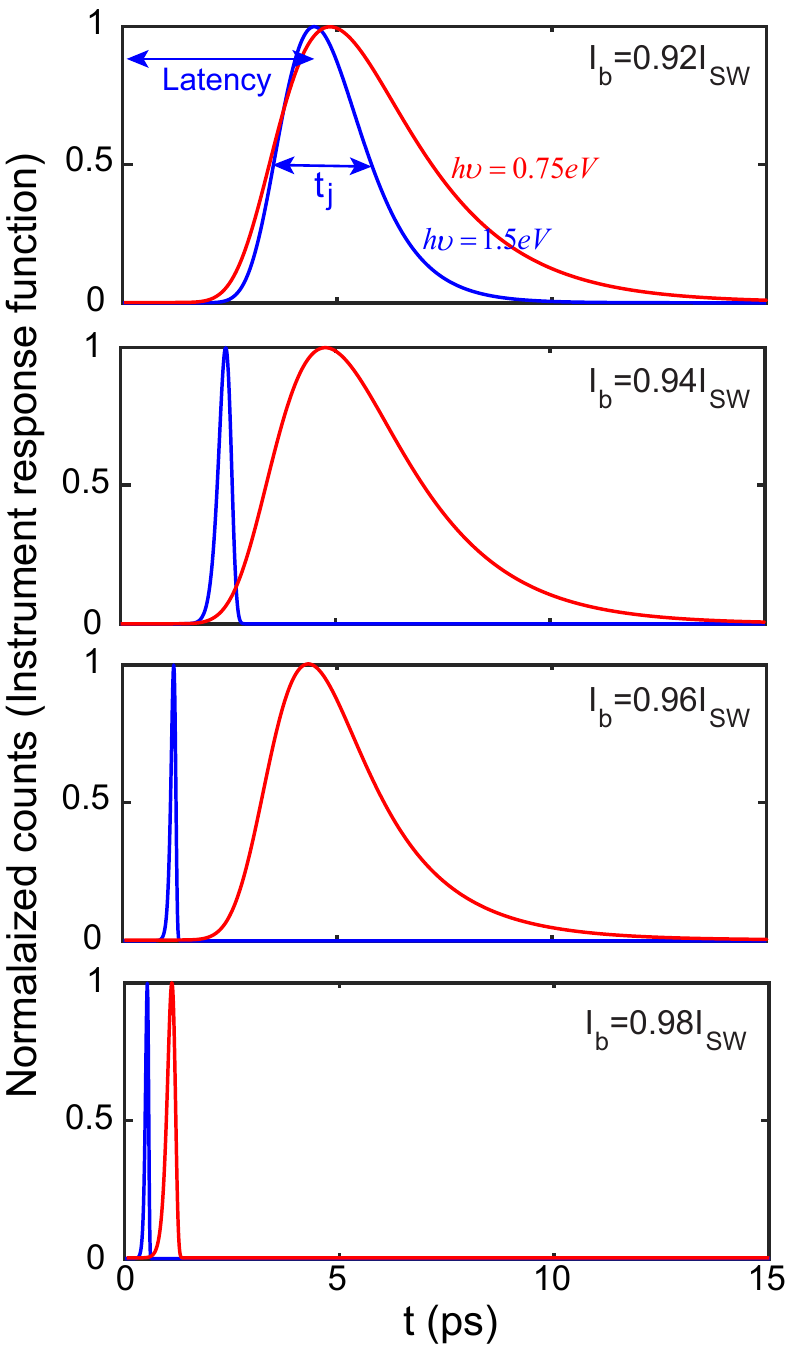}
\end{tabular}
\caption{ {\bf Instrument response function.} Estimated distribution of number of counts registered on the detector as a function of the delay after the photon absorption in a TaN SNSPD when the incoming single-photon energy is $h\nu=1.5$~eV (blue) and $h\nu=0.75$~eV (red); $T=0.6$~$^{\circ}$K and $W=100$~nm. There is a latency between the photon absorption and the click registration and the uncertainty in the latency causes a timing jitter ($t_j$) in the detector. 
}
\label{fig:SSPD_IRF}
\end{figure}

\begin{figure}[htbp]
\centering
\begin{tabular}{cc}
\includegraphics[width=8.5cm]{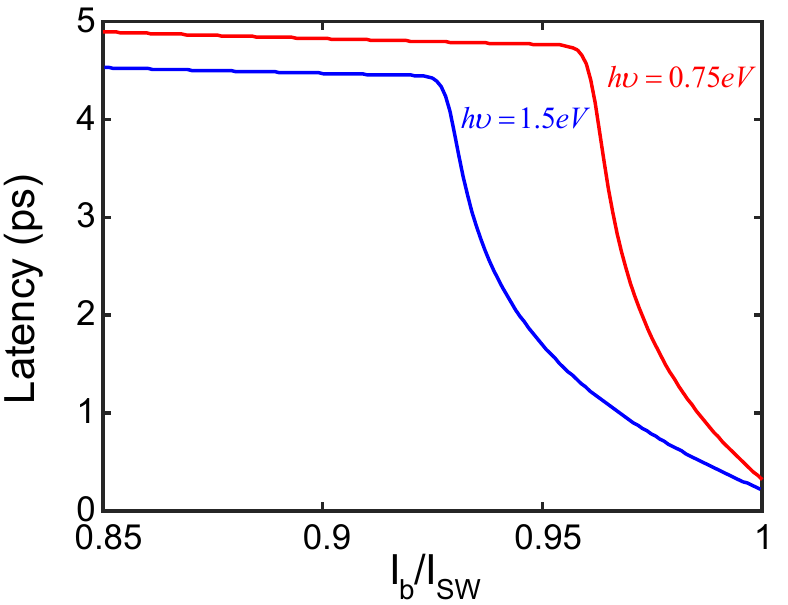}
\end{tabular}
\caption{ {\bf The effect of bias current on the latency in TaN SNSPDs.}  The latency suddenly drops when the quantum efficiency approaches unity, and it can be significantly reduced if the bias current becomes very close to the switching current ($I_{SW}$) or the photon energy increases. $T=0.6$~$^{\circ}$K and $W=100$~nm.
}
\label{fig:SSPD_lATENCY}
\end{figure}

\section{Quantum timing jitter}

According to the most accepted quantum measurement theory, positive-operator-valued measure (POVM), a quantum detector can be regarded as a black box. Each of its outcomes is represented by a positive Hermitian operator $\hat{\Pi}_m$ with non-negative real eigenvalues. The probability that the $m$th outcome occurs in experiments is given by $P_m=\rm{Tr}[\rho \hat{\Pi}_m]$, where $\rho$ is the initial state of the quantum object to be detected, such as the state of the incident single-photon pulse. The completeness condition, $\sum_m \hat{\Pi}_m = \hat{I}$ ($\hat{I}$ is the identity operator), expresses the fact that the probabilities sum to one: $\sum_m P_m =1$. For a non-photon-number-resolving photon detector, there are only two possible outcomes: clicking and non-clicking, characterized by $\hat{\Pi}_c$ and $\hat{\Pi}_{nc}$ (thus $m=c,nc$), respectively. The clicking probability, $P_c = \rm{Tr}[\rho \hat{\Pi}_c] \equiv$ $P_1 + P_0$, contains two parts: the single-photon induced clicking probability, $P_1$, characterizing the quantum efficiency of the detector and the dark counting part, $P_0$. Note that we have neglected the nonlinearity in the detector response \cite{akhlaghi2009semiempirical, akhlaghi2011nonlinearity}. Recently, the figures of merit and time-dependent spectrum of a single photon in terms of POVMs have been exploited~\cite{van2017photodetector,van2018time}. In the following, we present our microscopic calculation of $P_1$ and $P_0$ based on the single-vortex crossing model. Especially, we introduce the quantum timing jitter in the amplification process, which has not been incorporated into current POVM theory.

\begin{figure}[htbp]
\centering
\begin{tabular}{cc}
\includegraphics[width=8.5cm]{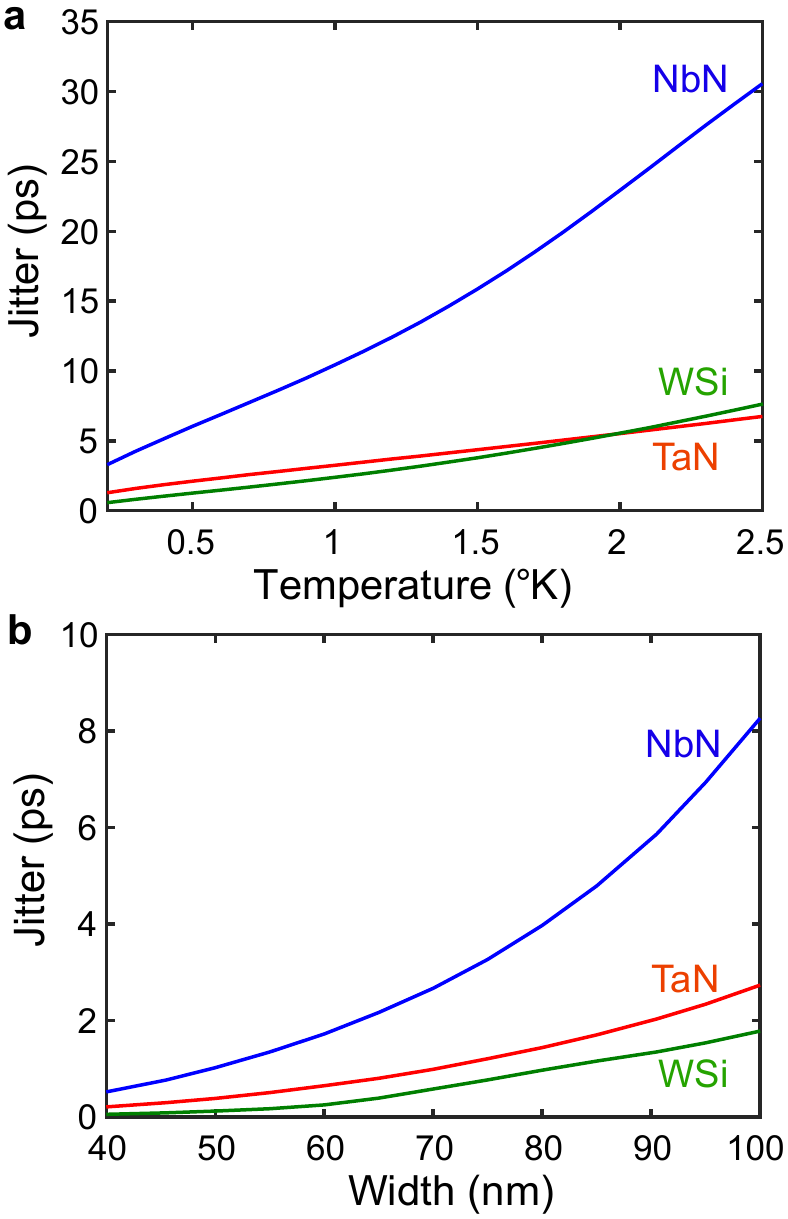}
\end{tabular}
\caption{ {\bf Timing jitter corresponding to vortex crossing.} Timing jitter in NbN, TaN, and WSi SNSPDs as a function of {\bf (a)} temperature ($W=100$~nm) and {\bf (b)} nanowire width ($T=0.6$~$^{\circ}$K. $h\nu=1.5$~eV. Decreasing the temperature results in a sharper change in the vortex crossing rate. Hence, the uncertainty of vortex crossing reduces. Reducing the width of the nanowire causes the QPs to distribute faster across the width of the nanowire, and as a result, the potential barrier reduces rapidly. 
}
\label{fig:SSPD_Jitter}
\end{figure}

Even if there is no photon and the bias current is below the vortex critical current $I_{c,v}$, a vortex can be thermally excited and escape the potential barrier saddle point to form a normal conducting belt \cite{bartolf_current-assisted_2010, eftekharian2013quantum}. This false-count rate is known as dark-count rate which deteriorates the performance of a single-photon detector \cite{hadfield_single-photon_2009}. The time-dependent rate of the vortex crossing can be described as: 
\begin{equation}\label{eq:Gamma}
\Gamma_{v}(t) =\alpha_{v} I_{b} \exp ({-U_{v,\max }(t) \mathord{\left/ {\vphantom {-U_{v,\max }  k_{B} T}} \right. \kern-\nulldelimiterspace} k_{B} T}),
\end{equation} 
where $k_B$ is the Boltzmann constant and $\alpha_v$ is a constant which is measured experimentally \cite{bartolf_current-assisted_2010}. $U_{v,\max }(t)$ is the maximum of the potential barrier for vortex crossing which changes with time after the single photon absorption event. As a result of the change in the vortex potential barrier after the photon absorption, the vortex crossing rate  increases exponentially. Figure~\ref{fig:SSPD_Gamma}a shows the vortex crossing rate as a function of time after the photon absorption for three different materials. The rate at $t=0$ corresponds to the dark-count rate \cite{yamashita_origin_2011,bartolf_current-assisted_2010, clem_geometry-dependent_2011,murphy_three_2015}. However, the rate is enhanced several orders of magnitude when the potential barrier reaches its minimum. This enhancement during the multiplication and recombination of QPs might be enough to significantly change the probability of vortex escaping the barrier. It is seen that the shape of the vortex crossing rate is different in different superconducting materials depending on the number of QPs generated and how fast they get multiplied, diffused across the width, and recombined. Since the crossing of vortices occurs independent of the other vortices, the crossing events can be regarded as a Poisson process with distribution function,
\begin{equation}
p(n_v,t)=\frac{\bar{n}_v(t)}{n_v!}e^{-\bar{n}_v(t)},
\end{equation}
characterizing the probability of $n_v$ vortex crossing the nanowire during the time interval $[t_0,t]$. Here, the time-dependent function $\bar{n}_v(t)=\int_{t_0}^t \Gamma_v(t')dt'$ is the mean number of vortices crossing the nanowire. Hence, we can define the single-photon detection probability $P_1$ after the single-photon absorption ($t_0=0$) and before time $t$ as the probability of crossing of at least one vortex as:
\begin{equation}\label{P1} 
P_{1} (t)=1-p(0,t)=1-\exp\left[-{\int _{t_0=0}^{t}\Gamma _{v} \left(t'\right)dt'}\right].
\end{equation}
We have neglected the interaction between vortices during the vortex crossing.

\begin{figure}[htbp]
\centering
\begin{tabular}{cc}
\includegraphics[width=8.5cm]{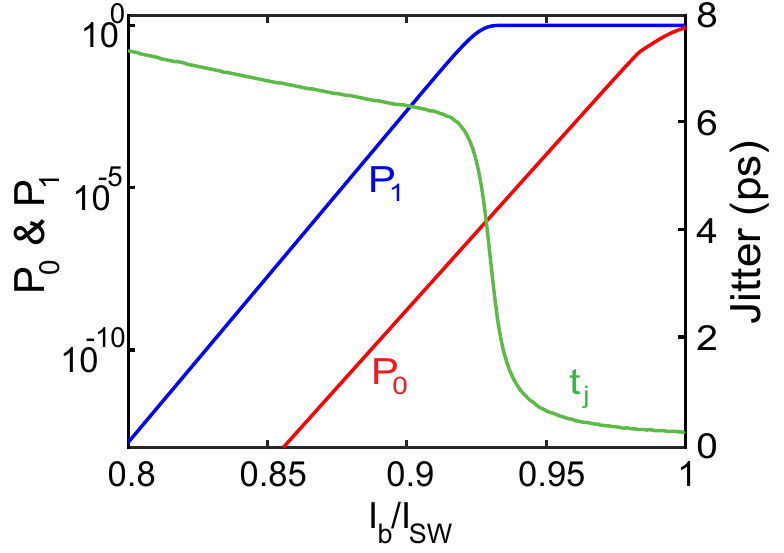}
\end{tabular}
\caption{ {\bf The effect of bias current on TaN SNSPD performance.} {\bf a,} Timing jitter, dark count probability ($P_0$), quantum efficiency ($P_1$) versus the bias current ($I_{b}$). $I_{b}$ is normalized to the switching current. Note that $I_{\rm SW}$ is smaller than $I_{c,v}$. Increasing the bias current helps to improve the detection probability and the timing jitter but at the cost of an increase in the dark count probability. $T=0.6$~$^{\circ}$K, $W=100$~nm, and $h\nu=1.5$~eV.}
\label{fig:SSPD_Bias}
\end{figure}
 
As seen in Fig.~\ref{fig:SSPD_Gamma}b, the detection probability increases rapidly around the crossing rate maximum. The time derivative of $P_1(t)$ is proportional to the single-photon count rate (also known as instrument response function) measured in experiments \cite{korzh_demonstrating_2018}. If the detection efficiency is low, the photon count rate is approximately the same as $\Gamma_v(t)$. The rise time of the quantum efficiency is not instantaneous due to the finite diffusion speed of the QPs and the hot-spot formation. Hence, there is a fundamental latency and uncertainty between the time of photon absorption and the quantum vortex tunneling process \cite{zhang2003time}. This causes a quantum timing jitter ($t_j$) on photon detection event as illustrated in Fig.~\ref{fig:SSPD_IRF}. As shown in Fig.~\ref{fig:SSPD_lATENCY}, the latency is lower for higher energy photon detection since the vortex barrier is suppressed faster. The latency is also reduced if the bias current is increased. There is a sharp drop in the latency when the photon detection becomes deterministic.

This type of timing jitter is because of the probabilistic tunneling of vortices \cite{yang_concept_2018, yang2019quantum}, which is nonzero even if the absorption location of the photon is known exactly. In the state-of-the-art experiments, the total jitter is dominated by the geometric jitter as a result of the uncertainty in the position of the transduction event \cite{wu_vortex-crossing-induced_2017, calandri_superconducting_2016, oconnor_spatial_2011, cheng_inhomogeneity-induced_2017, allmaras2019intrinsic, vodolazov2019minimal, kuzmin2019geometrical}. However, even if the geometric jitter is suppressed by defining the exact longitudinal \cite{zhao_single-photon_2017,cheng2019broadband} and transverse location of the photon absorption, the quantum jitter cannot be diminished beyond a limit. However, it can be controlled by engineering the structure and controlling the experimental conditions. Note that to make a quantitative comparison between the simulation results and the existing experimental results, the polarization of incident photon and the absorption location in the transverse co-ordinate must be considered \cite{renema2015position}. This will be undertaken in a future study.

\begin{figure}[ht]
\centering
\begin{tabular}{cc}
\includegraphics[width=8.5cm]{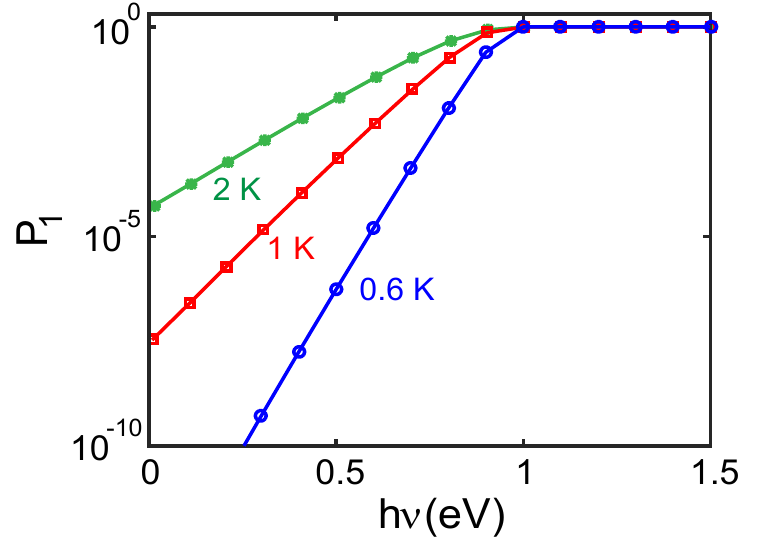} 
\end{tabular}
\caption{ {\bf Single-photon detection probability (quantum efficiency) as a function of single-photon energy and temperature in TaN SNSPDs}. The bias current is set to have near unity detection probability when the photon energy is larger than 1 eV. $W=100$~nm. For high energy photons, the vortex potential barrier drops to zero. Hence, the click event which is as a result of the vortex crossing happens certainly. However, if the photon energy is not high enough to suppress the potential barrier completely, the vortex crossing event becomes non-deterministic, and it drops exponentially as the photon energy is reduced. 
}
\label{fig:SSPD_Quantum_Efficiency}
\end{figure}

It is seen in Eq. (\ref{eq:Gamma}) that as the temperature decreases, the change in the vortex crossing rate becomes sharper. This causes the intrinsic timing jitter to reduce as shown in Fig.~\ref{fig:SSPD_Jitter}a which is in agreement with recent experiments \cite{santavicca2019jitter}. Note that the vortex potential is proportional to the characteristic vortex energy, $\varepsilon_0$. Thus, smaller vortex energy in causes a slower change in the vortex crossing rate, similar to the effect of rising the temperature. Hence, although the hot-spot formation and relaxation happens faster in WSi due to the smaller bandgap and faster QP diffusion \cite{engel_detection_2015}, the timing jitter in WSi is comparable with that in TaN because of the smaller $\varepsilon_0$ in WSi nanowires.

\begin{figure*}
\centering
\begin{tabular}{cc}
\includegraphics[width=14.25cm]{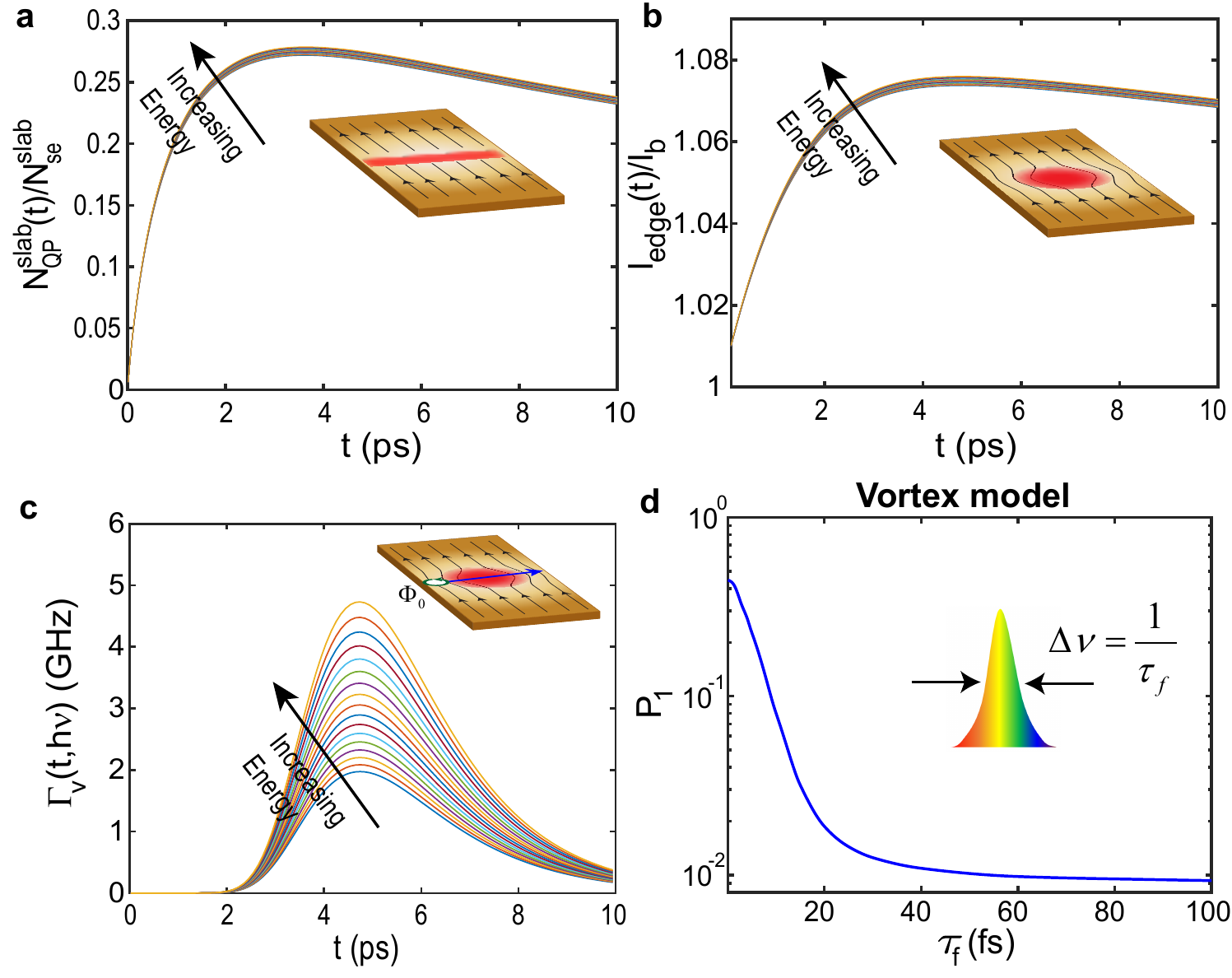}
\end{tabular}
\caption{ {\bf Energy-dependence and pulse-width dependence of the quantum efficiency.} {\bf a,} Normalized QP numbers inside the $\xi$-slab, {\bf b,} normalized current density at the edge, and, {\bf c,} vortex crossing rate for different modes of a photon pulse with the central energy of $h\nu=0.75$~eV and a pulse-width of $\tau_f=100$~fs; $T=0.6$~$^{\circ}$K and $W=100$~nm. The insets illustrate the schematic and the detection criteria for each model. It is seen that the number of QPs and current at the edge is not very sensitive to the small change of the photon energy. However, the single-vortex crossing rate is extremely sensitive as a result of a few percent change in the photon energy. {\bf d,} Single-photon quantum efficiency versus the pulse-width in vortex model. The quantum efficiency is considerably increased for the very short pulses. This experimental test can verify the validity of our model.
}
\label{fig:SSPD_Comparison}
\end{figure*}

Reducing the width of the nanowire results in a faster distribution of the hot-spot across the nanowire width. This leads to a sudden change in the vortex potential barrier, and as a result, the timing jitter decreases considerably as shown in Fig.~\ref{fig:SSPD_Jitter}b. Reducing the width helps reducing the geometrical timing jitter as well \cite{wu_vortex-crossing-induced_2017}, however, at the cost of a decrease in the transduction efficiency of the device. 

Increasing the bias current reduces the vortex potential and increases the vortex crossing rate. This causes not only an increase in the quantum efficiency ($P_1$) \cite{renema2013universal, caloz_optically_2017}, but also an increase in the dark count probability ($P_0$) as shown in Fig.~\ref{fig:SSPD_Bias}. $P_0$ is defined as the probability of the click while there is no interaction between the photon and the detector in the time bin of the photon arrival. $I_{\rm SW}$ is defined as the minimum bias current which is required for at least one vortex to escape the barrier in the time bin of the photon arrival. Note that $I_{\rm SW}$ is lower than the vortex critical current, $I_{c,v}$, especially if the temperature is not low enough. The bias current has also a significant impact on the timing jitter corresponding to single-vortex crossing. If the detector is biased very close to the switching current, a small perturbation due to the single-photon absorption suppresses the potential barrier and vortex can cross the width. Figure~\ref{fig:SSPD_Bias} displays the effect of the bias current on the timing jitter as well. It is seen that the timing jitter drops remarkably when the quantum efficiency approaches unity in agreement with the recent experimental observations \cite{sidorova2017physical, korzh2018wsi, caloz2018high}. This is because of the significant suppression of the barrier which leads to the vortex crossing even before the rate reaches its maximum. It is also seen that the slopes of $P_0$ and $P_1$ are identical in non-deterministic region in agreement with the recent experiments \cite{korneeva2018optical, knehr2019nanowire}, which confirms the connection between the photon counts and the dark counts around the detection current.

\section{Spectral quantum efficiency}

The non-deterministic behavior of vortices in the case when the potential barrier has not vanished completely allows us to estimate the quantum efficiency probability even for low energy photons. 
Figure~\ref{fig:SSPD_Quantum_Efficiency} shows the quantum efficiency based on the single-vortex crossing model as a function of the energy of the absorbed single-photon at different temperatures. The bias current is set to have a near unity quantum efficiency when the photon energy is larger than 1 eV. $\alpha_v$ in Eq.~(\ref{eq:Gamma}) can be used as a fitting parameter to define the actual current with respect to the experimental depairing current. The quantum efficiency approaches $P_0$ when the photon energy goes to zero. As the photon energy goes up, more changes in QP and current distributions are observed. This causes further suppression of the vortex potential barrier leading to a higher probability of the vortex crossing. This in turn results in a higher quantum efficiency.  Our vortex model by itself can explain both constant efficiency at high energies and exponential decrease of quantum efficiency at lower energies seen in experiments \cite{semenov_vortex-based_2008, hofherr2010intrinsic}.
Note that the photon absorption efficiency is assumed to be one over the entire spectrum. In practice, the absorption efficiency of a bare nanowire is not very high and does not vary significantly at optical frequencies. However, to increase the absorption efficiency, the detector should be placed inside a high-Q cavity \cite{rosfjord_nanowire_2006, vetter2016cavity, munzberg_superconducting_2018, erotokritou2018nano, yun2019superconducting} or a low-mode size waveguide \cite{ferrariwaveguide, najafi_-chip_2015, marsili_detecting_2013, jahani_transparent_2014, jahani_controlling_2018, akhlaghi_waveguide_2015}
to enhance the spatial overlap between the optical mode of the incoming photon and the superconducting electrons of the detector.

Till now, we have assumed in our model that the incoming photon is a single-mode photon. However, in practice, the photon has a finite pulse-width and the bandwidth of the photon may affect the performance of a detector. The response of the detector to a broadband photon can be used as an experimental test to compare across different detection models and verify our theory. In Fig.~\ref{fig:SSPD_Comparison}a-\ref{fig:SSPD_Comparison}c, we have compared the detection criteria in different models in response to the different modes of a multi-mode single-photon pulse with central energy of $h\nu=0.75$~eV and a pulse width of $\tau_f=100$~fs. Number of QPs and current at the edge are the main quantities to define detection criteria in hot-spot model and QP model \cite{engel_numerical_2013}. As shown in Fig.~\ref{fig:SSPD_Comparison}a and \ref{fig:SSPD_Comparison}b, the detector performance is not very sensitive when the photon energy is slightly changed around the central frequency of the photon. However, as shown in Fig.~\ref{fig:SSPD_Comparison}c, a small perturbation in the photon energy can make a considerable change in the single-vortex crossing rate since the rate exponentially changes with the vortex potential energy. Figure~\ref{fig:SSPD_Comparison}d displays the effect of pulse-width on the quantum efficiency in our model. It is seen that there is a remarkable change in the quantum efficiency for ultra-short single-photon pulses. This effect arises due to the exponential tail of the quantum efficiency and clearly differentiates the proposed vortex model from the existing detection mechanisms. A controlled experiment can verify whether our model is correct or not. 

\section{Conclusion}

In summary, we have proposed a probabilistic detection criterion in SNSPDs based on a single-vortex moving across the width of the detector. We have shown that even for a non-vanishing vortex potential barrier, there is a significant enhancement in the rate of the vortex crossing after the photon absorption leading to an increase in the click probability. This non-deterministic process insets a considerable intrinsic timing jitter to the detection event. We have shown the trade-space of the timing jitter, quantum efficiency, and dark counts for different superconducting materials and different nanowire structures. We have presented the quantum efficiency spectrum based on our model, which can predict a pulse-shaped dependent quantum efficiency in SNSPDs. This effect is negligible in other proposed models. Our model can predict some observables illustrated in Table~\ref{observables1} which can be verified experimentally to confirm or reject our model. It applies specifically to probabilistic (quantum) sources of jitter and further experiments are needed to distinguish such sources from the dominant geometric jitter.

\begin{table}
\centering
\caption{{\bf Experimental tests and observables to verify the validity of a detection model.}}
\label{observables1}
\begin{tabular}{|c|c|c|}
\hline
\multicolumn{1}{|c|}{\textbf{\shortstack{Experimental test}}}       & \textbf{\shortstack{Observables}}\\ \hline
Quantum efficiency spectrum         & \begin{tabular}{@{}c@{}}Exponential decrease at \\ low energies\end{tabular}  \\ \hline
Response function        			& \begin{tabular}{@{}c@{}}Latency vs. bias current  and\\  photon energy and\\ shoulder at threshold\end{tabular}  \\ \hline
Timing jitter vs. bias current	    & Shoulder at threshold  \\ \hline
Broadband single photon             & \begin{tabular}{@{}c@{}}Pulse-width dependency of \\ quantum efficiency\end{tabular}  \\ \hline
\end{tabular}
\end{table}

\appendix
\section{Detection mechanism formalism}

To find the time-dependent current and QP distributions, we use a modified semi-classical diffusion model which has been originally proposed by Semenov et. al \cite{semenov_quantum_2001} and developed by Engel and Schilling \cite{engel_numerical_2013,engel_detection_2015}.

\subsection{Quasi-particle multiplication}

 We assume the photon energy ($h\nu $) is considerably larger than the superconducting bandgap ($\Delta$), yet not large enough to make a phase transition and form a normal conducting core at the position of the photon absorption. Hence, when the photon is absorbed, a hot electron with a probability density of  $C_{e} (\vec{r},t)$ is created. Since the photon energy is usually orders of magnitude larger than the bandgap, when the hot electron diffuses, it breaks a large number of Cooper pairs ($> 100$ in the visible range) to QPs with a distribution density of $C_{qp} (\vec{r},t)$. This causes the hot electrons to lose their energy, and as a result, the multiplication process slows down with a life-time of $\tau _{qp}$ due to  electron-phonon interaction \cite{engel_numerical_2013}:
\begin{align}\label{eq:EqnSSPD1}
\frac{\partial C_{e} (\vec{r},t)}{\partial t} &= D_{e} \nabla ^{2} C_{e} (\vec{r},t),\\
\frac{\partial C_{qp} (\vec{r},t)}{\partial t} &= D_{qp} \nabla ^{2} C_{qp} (\vec{r},t)-\frac{C_{qp} (\vec{r},t)}{\tau _{r} }  \nonumber \\ &+ \frac{\varsigma h\nu }{\Delta \tau _{qp} } \left(\frac{n_{se,0} -C_{qp} (\vec{r},t)}{n_{se,0} } \right)e^{-{t\mathord{\left/ {\vphantom {t \tau _{qp} }} \right. \kern-\nulldelimiterspace} \tau _{qp} } } C_{e} (\vec{r},t), 
\end{align}
where $D_{e}$, $D_{qp}$, $\tau _{r}$, and $n_{se,0}$ are the hot-electron diffusion coefficient, quasi-particle diffusion coefficient, recombination time, and density of superconducting electrons before the photon absorption, respectively. $\varsigma$ is the QP conversion efficiency which has been assumed constant. We add the term ${\left(n_{se,0} -C_{qp} (\vec{r},t)\right)\mathord{\left/ {\vphantom {\left(n_{se,0} -C_{qp} (\vec{r},t)\right) n_{se,0} }} \right. \kern-\nulldelimiterspace} n_{se,0}}$ to include the saturation of QP multiplication. We have ignored the electron-phonon and phonon-phonon interactions which are considerably slower than the electron-electron interactions \cite{engel_numerical_2013, gousev1994broadband}.
The exact solution of the above equation in a general form is not easy to derive. Hence, to find the solution numerically, we have used a Finite-Difference Crank-Nicolson method. Since, the hot-electrons diffuse quickly ($D_{e} \gg D_{qp} $), to speed-up the simulations, we have used the analytical solution of Eq.~\eqref{eq:EqnSSPD1} for the case of an infinite 2D superconductor \cite{engel_numerical_2013}. We have assumed a Gaussian distribution for the electron, which is a delta function at $t=0$ and the electron diffuses for $t>0$. A grid size of $\Delta x=\Delta y=1-3$~nm and a time step of ${D_{qp} \Delta t\mathord{\left/ {\vphantom {D_{qp} \Delta t \Delta x^{2} }} \right. \kern-\nulldelimiterspace} \Delta x^{2} } =0.01$ is used in our simulations. Neumann boundary condition for the side-walls and zero-flux at the two ends of the nanowire have been considered.  The material parameters can be derived from experimental measurements \cite{engel_detection_2015, bartolf_current-assisted_2010}. The parameters that we have used in this work are listed in Table~\ref{tab:material}.

\begin{table}
    \centering
\caption{{\bf Material parameters near zero temperature used in simulations}}
\begin{tabular}{l*{7}{c}r}

              & $\Delta$ & $D_e$ & $D_{qp}$ & $\xi$ & $\lambda$  & $\tau_r$ &  $\tau_{qp}$ & $\varsigma$\\

              & (eV) & (nm$^2/$ps) & (nm$^2/$ps) & (nm) & (nm)  & (ps) &  (ps) & (\%)\\
\hline
\hline
TaN            & 1.3 & 8.2 & 60 & 5.3 & 520 & 1000 & 1.6  & 25 \\
NbN            & 2.3 & 7.1 & 52 & 4.3 &  430 & 1000 &  1.6  &  25\\
WSi            & 0.53 & 10.3 & 75 & 8 &  1400 & 1000 &  1.6 &  25  \\
\end{tabular}
    \label{tab:material}
\end{table}

\subsection{Current redistribution}

The current distribution can be calculated by combining superconducting phase coherence condition and continuity equation \cite{tinkham_introduction_1996}:  
\begin{equation}
\nabla .(\vec{j}(\vec{r},t))=\nabla .\left(\frac{\hbar }{m} n_{se} (\vec{r},t)\nabla \varphi(\vec{r},t) \right)=0,
\end{equation}
where $n_{se} (\vec{r},t)=n_{se0} -C_{qp} (\vec{r},t)$ is the density of superconducting electrons after the photon absorption, $\varphi$ is the phase of the superconducting order parameter, $m$ and $\hbar$ are the electron mass and reduced Planck constant, respectively.  

\subsection{Single vortex crossing}

Vortices and antivortices are the topological defects in thin superconducting films which exist even if there is no applied magnetic field \cite{tinkham_introduction_1996}. Vortices are usually nucleated and enter into the nanowire from the edge where the superconducting order parameter is suppressed. London equation in the presence of a static vortex in a superconducting thin film in $xy$ plane can be written as \cite{tinkham_introduction_1996, kogan_interaction_2007}:
\begin{equation}\label{EqnSSPD4}
\vec{H}(r)+2\pi \frac{\Lambda }{c} \nabla \times \vec{j}(r)=\hat{z}\Phi _{0} \delta \left(\vec{r}-\vec{r}_{v} \right), 
\end{equation}
where $\Lambda = 2\lambda^2/d $ is the Pearl length \cite{pearl_current_1964}, $\lambda $ is the London penetration depth, $d$ is the film thickness, $\Phi_0 =hc/2e$ is the magnetic flux quantum due to the presence of a single-vortex at the position $\vec{r}_{v} $, $\vec{H}$ is the magnetic field, $\vec{j}$ is the current density ignoring the effect of the vortex on the current, and $c$ is the speed of light in vacuum. Since the thickness of the nanowire is significantly smaller than $\lambda$, we have averaged the field and the current in the $z$ direction. For nanoscale SNSPDs ($L\ll\Lambda$), the first term can be neglected \cite{kogan_interaction_2007}, and because of the current continuity ($\nabla .\vec{j}=0$), we can write the current density in the form of a scaler function as $\vec{j}(r)=\nabla \times G(r)\hat{z}$. Thus, eqn.~\eqref{EqnSSPD4} is reduced to \cite{kogan_interaction_2007}:
\begin{equation}\label{eq:Poisson}
\nabla ^{2} G(r)=-\frac{c\Phi _{0} }{2\pi \Lambda } \delta \left(\vec{r}-\vec{r}_{v} \right),
\end{equation}
which is equivalent to the 2D Poisson's equation for a charged particle. For an infinite superconducting film case, the interaction energy between vortices and antivortices for distances shorter than the Pearl length is logarithmic. This allows Berezinskii-Thouless-Kosterlitz (BKT) transition  and the formation of vortex-antivortex pairs below the BKT critical temperature \cite{kadin_photon-assisted_1990, kogan_interaction_2007}. However, for a thin superconductor with finite width (${-W\mathord{\left/ {\vphantom {-W 2}} \right. \kern-\nulldelimiterspace} 2} <x<{W\mathord{\left/ {\vphantom {W 2}} \right. \kern-\nulldelimiterspace} 2} $), the long range interaction between vortices and antivortices is eliminated and single vortices can be found. For a single vortex, eqn.~(\ref{eq:Poisson}) is reduced to the equation for a charge sandwiched between two parallel grounded plates. The problem is well-known in electrostatics and can be solved using conformal mapping with $z'=e^{i{\pi z\mathord{\left/ {\vphantom {\pi z W}} \right. \kern-\nulldelimiterspace} W} }$ transformation and using image theory \cite{kogan_interaction_2007}:
\begin{equation}
G(x,y)=\frac{c\Phi _{0} }{8\pi \Lambda } \ln \frac{\cosh \left({y\pi \mathord{\left/ {\vphantom {y\pi  W}} \right. \kern-\nulldelimiterspace} W} \right)+\cos \left({(x+x_{v} )\pi \mathord{\left/ {\vphantom {(x+x_{v} )\pi  W}} \right. \kern-\nulldelimiterspace} W} \right)}{\cosh \left({y\pi \mathord{\left/ {\vphantom {y\pi  W}} \right. \kern-\nulldelimiterspace} W} \right)-\cos \left({(x-x_{v} )\pi \mathord{\left/ {\vphantom {(x-x_{v} )\pi  W}} \right. \kern-\nulldelimiterspace} W} \right)},
\end{equation}
where we have assumed the vortex is placed at $x=x_{v} $ and $y=0$. The phase of the order parameter, $\varphi$, can also be derived from $G$ since the gradient of $\varphi$ is also proportional to the current \cite{bulaevskii_vortex-induced_2011}:
\begin{align}\label{phase}
\varphi \left(\vec{r},\vec{r}_{v} \right)=\tan ^{-1} \frac{\cos \left(\frac{\pi x}{W} \right)\sinh \left(\pi \frac{y-y_{v} }{W} \right)}{\sin \left(\frac{\pi x}{W} \right)-\cosh \left(\pi \frac{y-y_{v} }{W} \right)\sin \left(\frac{\pi x_{v} }{W} \right)}. 
\end{align}

The free energy in presence of a vortex consists of the field energy and the kinetic energy inside the nanowire and the field energy outside \cite{tinkham_introduction_1996, kogan_interaction_2007}. If we assume the vortex core radius is $\xi$ and we neglect the core energy of the vortex, the self-energy of the vortex can be written as \cite{bulaevskii_vortex-assisted_2012}:
\begin{align}
U_{v}^{0} (x_{v} ) &=-\frac{\Phi _{0} }{2c} G(\left|x-x_{v} \right|\to \xi ,0) \nonumber \\ &=\frac{\Phi _{0}^{2} }{8\pi ^{2} \Lambda } \ln \left(\frac{2W}{\pi \xi } \cos \left(\frac{\pi x_{v} }{W} \right)\right), 
\end{align}

If we include the work done by the bias current on a single vortex due to the Magnus force (dual of the Lorentz force on a magnetic flux), the total energy of a single vortex is expressed as \cite{bulaevskii_vortex-induced_2011}:
\begin{align}
\!\!\!U_{v} (x_{v}) \!= \!\frac{\Phi _{0}^{2} }{8\pi ^{2} \Lambda } \ln \!\left(\! \frac{2W}{\pi \xi } \cos \left(\frac{\pi x_{v} }{W} \!\right)\! \right)\!- \!\frac{\Phi _{0} }{c} j_{y} (x_{v})(x_{v}+\frac{W}{2}).\!
\end{align}

The Magnus force tries to move the vortex in the direction perpendicular to the direction of the applied bias current, but it cannot overcome the self-energy of the vortex if the bias current is not high enough. Increasing the bias current at the edges $j_y(x_v,t)$ due to the photon absorption reduces the potential barrier and eases vortex crossing. This barrier finally turns to zero at the vortex critical current which is:
\begin{align}
I_{c,v} =\frac{c\Phi_{0}}{4\pi ^{2} \exp (1)\Lambda \xi }W. 
\end{align}

As seen in eqn.~(\ref{phase}), the phase of the order parameter depends on the position of the vortex, $x_v$, and the phase difference at the two ends of a long nanowire ($L \gg W$) away from the vortex position can be approximated as:
\begin{align}
\varphi(L/2)-\varphi(-L/2)=2\pi{x_v}/W.
\end{align}
Hence, as the vortex moves across the width of the nanowire, it applies a time-dependent phase difference between the two terminals of the detector. If the vortex crosses from one edge at $x_v=-W/2$ to another edge at $x_v=W/2$, it causes a $2\pi$ phase-slip at the two ends of the nanowire. This phase evolution generates a voltage pulse which can be described by the Josephson effect \cite{clem_time-dependent_1981}:
\begin{align}
V(t)=\frac{\Phi_0}{2\pi{c}}\frac{d}{dt}(\varphi(L/2)-\varphi(-L/2))=\frac{\Phi_0}{cW}\frac{dx_v}{dt}.
\end{align}

This voltage pulse propagates to the two ends \cite{santavicca_microwave_2016,zhao_single-photon_2017} and is dissipated in the presence of the bias current. If the bias current is high enough, the released energy is enough to induce a phase transition in the nanowire from the superconducting state to the normal conducting state. Hence, the current $I_{c,v}$, which makes the vortex tunneling barrier reduce to zero,  is also the critical current for phase transition in a thin superconducting nanowire. This critical current is less than the depairing critical current in a bulk superconductor, $I_{c,dep}$, \citep{renema_experimental_2014,engel_numerical_2013}.

Even if the applied bias current is below $I_{c,v}$, breaking of the Cooper pairs and redistribution of the bias current due to the photon absorption can also change the potential barrier \cite{engel_numerical_2013}:
\begin{align}
\frac{U_{v} (x_v,t)}{\varepsilon _{0} } &= \frac{\pi }{W} \int _{\frac{\xi -W}{2} }^{x_v}\frac{n_{se} (x',t)}{n_{se,0} } \tan \left(\frac{\pi x'}{W} \right)dx' \nonumber \\ &- \frac{2W}{I_{c,v} \exp (1)\xi } \int _{-\frac{W}{2} }^{x_v}\frac{n_{se} (x',t)}{n_{se,0} } j_{y} \left(x',t\right)dx', 
\end{align}
where $\varepsilon _{0}=\Phi_0^2/8\pi^2\Lambda $ is the characteristic vortex energy.

\section*{\label{sec:level1}Acknowledgement}

We thank Sean Molesky, Joseph Maciejko, Rudro Biswajs, and Bhaskaran Muralidharan for discussions. This work is supported by DARPA DETECT. 

\bibliography{SSPD}

\end{document}